\documentclass[twocolumn,floatfix,amsmath,amssymb,prl, superscriptaddress,10pt]{revtex4}

\usepackage{epsfig}
\usepackage{graphics}
\usepackage{dcolumn}
\usepackage{bm}
\usepackage{color}

\begin{document}

\title{Observation of a $p$-wave Optical Feshbach Resonance}
\author{Rekishu Yamazaki}
\altaffiliation[Currently at ]{Research Center for Advanced Science and Technology, The University of Tokyo, 4-6-1 Komaba, Meguro-ku, Tokyo 153-8904, Japan}
\affiliation{Graduate School of Science, Kyoto University, Kitashirakawa Oiwake-cho, Sakyo-ku, Kyoto, 606-8502, Japan}
\affiliation{JST-CREST, 4-1-8 Honmachi, Kawaguchi, Saitama 331-0012, Japan}
\author{Shintaro Taie}
\author{Seiji Sugawa}
\affiliation{Graduate School of Science, Kyoto University, Kitashirakawa Oiwake-cho, Sakyo-ku, Kyoto, 606-8502, Japan}
\author{Katsunari Enomoto}
\affiliation{Department of Physics, University of Toyama, Toyama 930-8555, Japan}
\author{Yoshiro Takahashi}
\affiliation{Graduate School of Science, Kyoto University, Kitashirakawa Oiwake-cho, Sakyo-ku, Kyoto, 606-8502, Japan}
\affiliation{JST-CREST, 4-1-8 Honmachi, Kawaguchi, Saitama 331-0012, Japan}
\date{\today}

\begin{abstract}
We demonstrate a $p$-wave optical Feshbach resonance (OFR) using purely long-range molecular states of a fermionic isotope of ytterbium $^{171}$Yb, following the proposition made by K. Goyal \textit{et~al.} [Phys.~Rev.~A {\bf 82}, 062704 (2010)].
The $p$-wave OFR is clearly observed as a modification of a photoassociation rate for atomic ensembles at about 5~$\mu $K.
A scattering phase shift variation of $\delta \eta=0.022$~rad is observed with an atom loss rate coefficient $K=28.0\times$10$^{-12}$~cm$^3$/s.
\end{abstract}


\maketitle

Ultracold atoms have recently been utilized as a versatile test bench for the study of many-body physics.
In particular, the capability to tune a wide range of interatomic interaction with fine controllability offered by magnetic Feshbach resonances (MFRs) \cite{Kohler2006, Chin2010} is so powerful and has enabled numerous novel observations, including a Mott insulator of fermionic atoms in an optical lattice \cite{Jordens2008}, Bose-Einstein condensation(BEC)-Bardeen-Cooper-Schriefer(BCS) crossover \cite{Regal2004, Zwierlein2004, Zwierlein2005}, and  Efimov trimers\cite{Kraemer2006}.
For these studies, alkali atoms equipped with the MFR have been the main work horse.
\begin{figure}[t]
  \includegraphics[width=8cm]{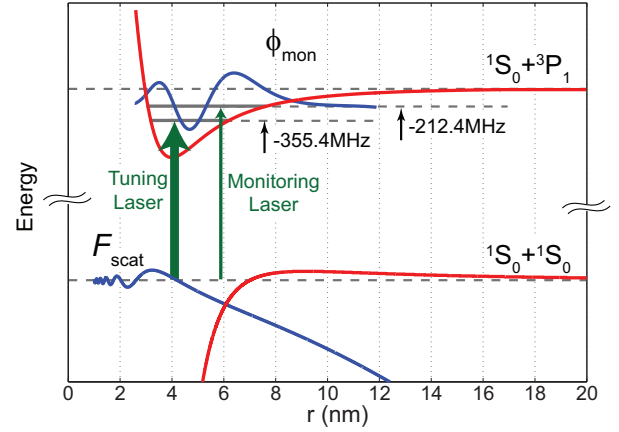}\\
  \caption{(color online) Relevant molecular states and adiabatic potentials for the $p$-wave OFR experiment.
  Tuning and monitor lasers ($\lambda=555.8$~nm) are tuned near PA resonances at $-355.4$~MHz and $-212.4$~MHz from the $^1S_0$+$^3P_1$ asymptote.   The excited states are previously observed PLR molecular states arising from the hyperfine interaction.  A part of the wavefunction of the $p$-wave scattering state $F_{\text{scat}}$ as well as the wavefunction of the state for the monitoring $\phi_{\text{mon}}$ are shown. Note that the energy scale of the ground state potential is enlarged by a factor of hundred compared with the excited state potential to show the centrifugal barrier of approximately 44~$\mu$K in height.} \label{fig1}
\end{figure}

\begin{figure*}[t]
  \includegraphics[width=17.7cm]{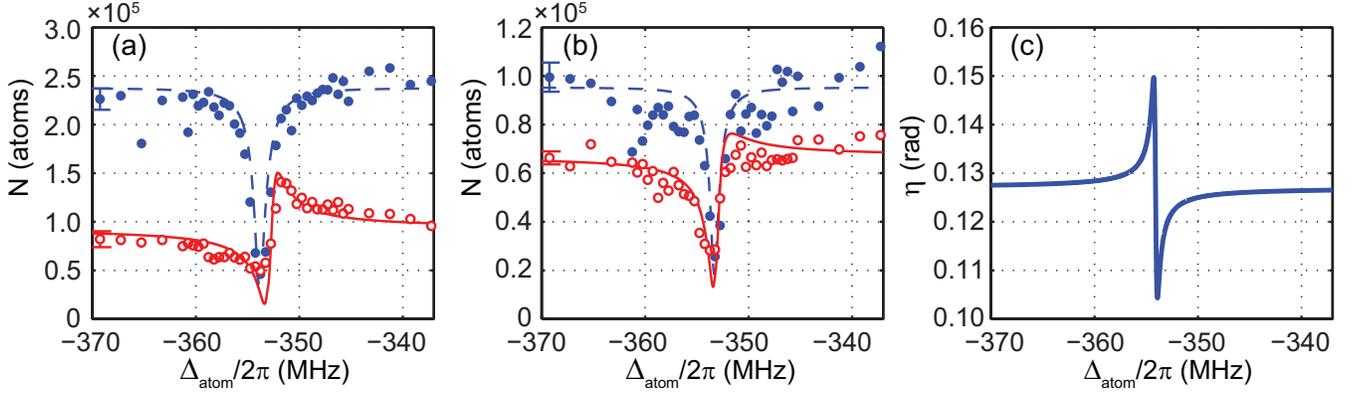}\\
  \caption{(color online) (a,b) $p$-wave OFR spectra with only the tuning laser (solid circles) and with both the tuning and monitoring lasers (open circles). Number of atoms remaining after the OFR pulse ($N$) is plotted with respect to the tuning laser detuning  from the $^1S_0$+$^3P_1$ asymptote ($\Delta_{atom}$).  The error bars in the figure shows the statistical uncertainty of number of atoms.  The data obtained with different temperatures, (a) $T=7.2$~$\mu$K and (b) 4.5~$\mu$K, are shown. The dotted lines show the atom loss spectra calculated with the laser induced linewidth $\Gamma_{\text{tun}}/2\pi=16.5$~kHz and 16.2~kHz, for (a) and (b), respectively.  The asymmetric atom loss spectra, indication of the OFR effect, is clearly observed with the monitor laser.  The spectra are well reproduced in solid lines with minimal fitting parameters as discussed in the main text.  (c) Calculated variation of $\eta$ using the value of $\Gamma_{\text{tun}}$ obtained from the fit for $T=7.2$~$\mu$K.  }\label{fig2}
\end{figure*}

An alternative approach, using an optical transition to artificially form a Feshbach resonance, \emph{optical Feshbach resonance} (OFR) was proposed \cite{Fedichev1996} and  successfully demonstrated using alkali atoms \cite{Fatemi2000, Theis2004, Wu2012}.
These experiments were accompanied with a rather large two-body inelastic atom loss due to the photoassociation (PA).
This loss can be mitigated, however, using a narrow intercombination transition in the alkaline-earth-like atoms, as suggested by Ciury\l{}o \textit{et~al.} \cite{Ciurylo2005}.
This narrow line OFR has been demonstrated experimentally in thermal gases and condensates of ytterbium(Yb) \cite{Enomoto2008, Yamazaki2010}, which showed about an order of magnitude suppression of the atom loss as compared to the case for the alkali atoms.
Later experiments using strontium atoms also showed similar advantages \cite{Blatt2011, Yan2012}.
Besides the suppression of an atom loss, the optical manner to control the interatomic interaction, including the optically controlled MFR \cite{Bauer2009}, introduces new possibilities in quantum gas manipulation such as a fast temporal manipulation and fine spatial resolution, which have been successfully demonstrated in Ref.\cite{Yamazaki2010}.
Arbitrary control of interatomic interactions among different kinds of atom pairs in a mixture of gases would be also an interesting possibility.
These technical advancements can broaden the spectrum of experiments that can be performed with quantum gases.

In all of the previous experiments, the effect of the OFR has been examined only on the $s$-wave interaction.
In this paper, we extend the ability of the OFR to control interatomic interaction of higher-partial waves \cite{Deb2009, Goyal2010}.
We successfully demonstrate the $p$-wave OFR effect in fermionic $^{171}$Yb atoms in the vicinity of the  purely long-range (PLR) molecular resonance
, the use of which has been suggested by Goyal \textit{et~al.} \cite{Goyal2010}.
An intriguing application of a $p$-wave OFR would be the study on the $p$-orbital bands of spinless fermions trapped in an optical lattice \cite{Goyal2010, Hauke2011}.  Such $p$-orbital physics in optical lattices has been discussed in Refs.~\cite{Wu2008, Wu2008a}.

The radial part of the energy-normalized ground-state $p$-wave scattering wavefunction has an asymptotic form
$F_{\text{scat}}(r,k)=(2\mu/\pi\hbar^2k)^{1/2}\sin(kr-\pi/2+\eta)$ at a long interatomic distance $r$,
where $\eta$ is the scattering phase shift, $\mu$ is the reduced mass of the scattering atom pair, and $k$ is the wave number of the atom pair, approximated as $k=\sqrt{2\mu k_B T }/\hbar$ with $T$ the temperature of the atomic sample, $k_B$ the Boltzmann constant, and $\hbar$ the Planck constant divided by $2\pi$.
The OFR effect is induced by the laser excitation to a molecular state, which alters the ground state scattering wavefunction.
Using the semi-analytic formalism by Bohn and Julienne \cite{Bohn1999},
a general form of the scattering matrix element $S=\exp(2i\eta)$ for the colliding ground state atoms modified by the OFR laser excitation is given by
\begin{equation}
S=\exp(2i\eta_0)\frac{\Delta-i\left(\Gamma-\gamma\right)/2}{\Delta+i\left(\Gamma+\gamma\right)/2},
\end{equation}
where $\eta_0$ is the scattering phase shift associated with the unperturbed scattering wavefunction.  $\gamma$ is the radiative decay rate of the excited molecular state, and $\Delta$ is the detuning of the OFR laser with respect to the molecular PA resonance.
$\Gamma$ is the laser-induced width given by
\begin{equation}\label{Gamma}
  \Gamma=\frac{\pi}{2}\left(\frac{I}{I_{sat}}\right)\hbar\gamma_a^2f_{FC},
\end{equation}
 where  $I$ is the OFR laser intensity, $I_{sat}=0.14$~mW/cm$^2$ and $\gamma_a/2\pi=182$~kHz are the saturation intensity and linewidth for the atomic $^3P_1$ state, respectively.  $f_{FC}=|\langle \phi_{\text{e}} | \mathbf{d} \cdot \mathbf{\epsilon}_L | F_{\text{scat}}\rangle|^2/2d_A^2$ is the Franck-Condon factor including the rotational correction, where $\mathbf{d}$, $d_A$, and $\mathbf{\epsilon}_L$ are the dipole moment operator, atomic dipole moment and the laser polarization, respectively. $\phi_{\text{e}}$ and $F_{\text{scat}}$ are the wavefunction of the excited state and scattering state, including the spin and rotational degree of freedom.  The variation of the scattering properties by the OFR effect can be determined from the $S$ matrix.
The loss rate coefficient $K$ and scattering phase shift $\eta$ can be derived from $S$-matrix as,
\begin{eqnarray}
K &=& \frac{\pi\hbar}{\mu k}(1-|S|^2)=\frac{\pi\hbar}{\mu k}\frac{\Gamma\gamma}{\Delta^2+\frac{(\Gamma+\gamma)^2}{4}} \label{K} \\
\eta &=& \arg(S)/2. \label{eta}
\end{eqnarray}

For the demonstration of the $p$-wave OFR effect, we tune the OFR laser near to the previously observed PLR molecular state arising from the hyperfine interaction \cite{Enomoto2008b}.  Relevant molecular states and adiabatic potentials are denoted in Fig.~\ref{fig1}.  The PLR molecular states have the inner turning point at $\sim$2.5~nm and have small Franck-Condon overlap with tightly bound states, possibly suppressing the inelastic atom loss.  In order to determine the variation of the scattering phase shift  in a systematic manner, we use the same technique previously demonstrated for the measurement of the $s$-wave OFR effect \cite{Fatemi2000, Enomoto2008}.
Along with the strong OFR laser (tuning laser) which tunes the scattering phase shift, we apply a much weaker laser (monitor laser) tuned right on another PA resonance.
PA resonances at $-212.4$~MHz and $-355.4$~MHz are associated with $T_{\text{e}}=3$ PLR states which are optically accessible from the $p$-wave scattering states, and are used for the monitor and tuning lasers, respectively, where $T_{\text{e}}$ is the total angular momentum of the molecular states.  Condon radius of the tuning and monitor states are $R_c=7.22$ and 5.96~nm, respectively.

In the previous demonstration of the $s$-wave OFR, the tuning laser varies the $s$-wave scattering phase shift, effectively translating the scattering wavefunction in the radial direction.
This translation results in the variation of the Franck-Condon factor for the monitor PA transition, which enables to observe the variation of the phase shift $\eta$ in terms of the monitor PA rate variation.  For the current study of the $p$-wave OFR, however, we should be careful due to the existence of the centrifugal barrier as discussed later in details.

The experimental setup is nearly the same as our previous works on the PA experiments \cite{Tojo2006, Kitagawa2008}.
A typical atom number obtained after the evaporation is $1.7 \times 10^5$ at 5.6~$\mu$K.
After the preparation of the cold atom sample with mixed spin, the tuning and monitor lasers with $\lambda=556$~nm are turned on to perform the OFR experiment.
The two beams are combined and the polarization is cleaned with a polarizer to the same polarization before sent into a polarization maintaining fiber for the experiment.  Linearly polarized beams are focused at the atomic cloud with the beam waist of 70~$\mu$m.
The tuning laser power is kept at 525~$\mu$W, which is the maximum power obtainable with the current setup, while monitor laser power is adjusted to 113~$\mu$W and 37~$\mu$W for $T=7.2$~$\mu$K and 4.5~$\mu$K, respectively.  Two beams are sent in together with the pulse duration of 30~ms.
The monitor laser frequency is fixed to the peak of the PA resonance at $-212.4$~MHz, while the tuning laser is scanned around the PA resonance at $-355.4$~MHz.
The atom loss induced by the two lasers is monitored by taking the absorption image of the sample.
We perform the measurements at a nearly zero magnetic field.  The residual magnetic field is estimated to be below 50~mG, which results in the Zeeman splitting of the excited state sublevels to less than 2~kHz, well within the linewidth of the excited state.

The observed atom loss spectra are shown in Fig.~\ref{fig2} for (a) $T=7.2$~$\mu$K and (b) 4.5~$\mu$K.
The $p$-wave centrifugal barrier for Yb atoms is approximately 920~kHz (44~$\mu$K), and
we were unable to observe clear PA spectra below 3.5~$\mu$K, due to lower starting atom number from the large loss in the evaporative cooling and PA rate suppression at lower temperature.
With the tuning laser only, the spectra are symmetric as expected for the ordinary PA resonance.  When the monitor laser is turned on, the additional atom loss due to the monitor laser PA is introduced, which is observed for the data at off-resonance in Fig.~2(a) and 2(b).
The OFR effect manifests itself around the tuning laser PA resonance, where the dispersive-shaped spectra are observed.  This behavior suggests that the Franck-Condon factor between the scattering wave state and the monitor state is altered dispersively around the tuning PA resonance, with the effect diminishing  at a lower temperature.

\begin{figure}[t]
  \includegraphics[width=8cm]{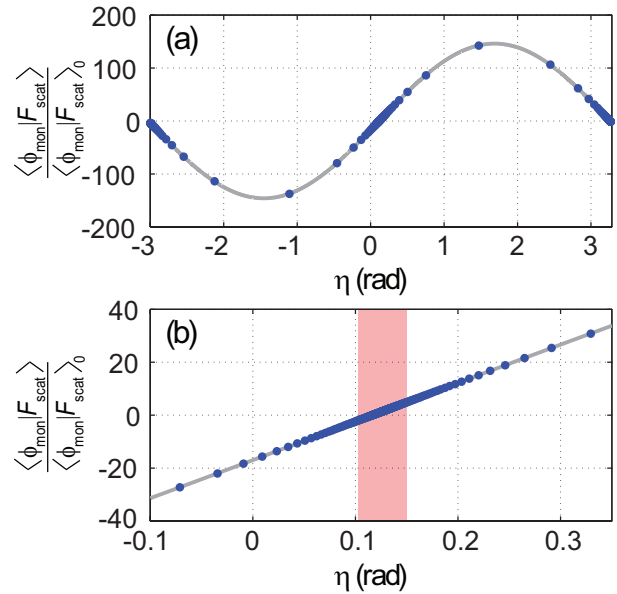}\\
  \caption{(color online) Calculated wavefunction overlap integral, $\langle\phi_{\text{mon}}|F_{\text{scat}}\rangle$, for the monitor state ($-212.4$~MHz) and the scattering state as a function of the phase shift $\eta$, for (a) a wide range and (b) near the background phase shift $\eta_0=0.127$ at $T=5.6$~$\mu$K.  The overlap integral is normalized with respect to the overlap integral at $\eta_0$, denoted as $\langle\phi_{\text{mon}}|F_{\text{scat}}\rangle_0$. Dots are the numerical results with lines showing the guide to the eye.  Large overlap integral can be seen near $\eta=1.7$ and $-1.4$, representing the location of the shape resonances.  Near $\eta_0$ the variation of the overlap integral is quite linear with respect to $\eta$, with zero at $\eta'=0.116$.  The shaded region shows the range where the variation of $\eta$ is observed at a current study.} \label{fig3}
\end{figure}

For the quantitative analysis of the OFR spectra with the monitor laser on, it is important to analyze the overlap integral between the scattering wavefunction of the ground state $F_{\text{scat}}$ and that of the molecular bound state $\phi_{\text{mon}}$ used for the monitoring PA.  We performed numerical calculation of the bound and scattering wavefunction using Numerov method with a potential provided by van der Waals potential with potential constant C$_6$ = 1931.7~a.u. and the centrifugal term.
The monitor-laser-induced atom loss rate coefficient $K_{\text{mon}}$ is proportional to $\Gamma_{\text{mon}}$ when the  laser is tuned on-resonance ($\Delta_{\text{mon}}=0$) with the weak excitation $\Gamma_{\text{mon}} \ll \gamma$, as understood from Eq.~\ref{K}.

As shown in Fig.~1, the wavefunction of the molecular bound state $\phi_{\text{mon}}$ is mostly inside the centrifugal barrier ($r<$7.0~nm) of the $p$-wave scattering state.
The calculated wavefunction overlap integral, $\langle\phi_{\text{mon}}|F_{\text{scat}}\rangle$ as a function of the scattering phase shift $\eta$ is shown in Fig.~\ref{fig3}(a).
The overlap integral varies sinusoidally with respect to $\eta$.
The background scattering phase shift calculated for $^{171}$Yb is $\eta_0=0.127$ at $T=5.6$~$\mu$K.
As shown in Fig.~\ref{fig3}(b), the variation of the overlap integral is quite linear in the vicinity of $\eta_0$, with zero overlap at $\eta'=0.116$.
Therefore, we expect the form of the monitor-laser-induced atom loss rate coefficient $K_{\text{mon}}=K_0(\eta-\eta')^2$ for the current experiment.

 \begin{figure}[t]
  \includegraphics[width=7cm]{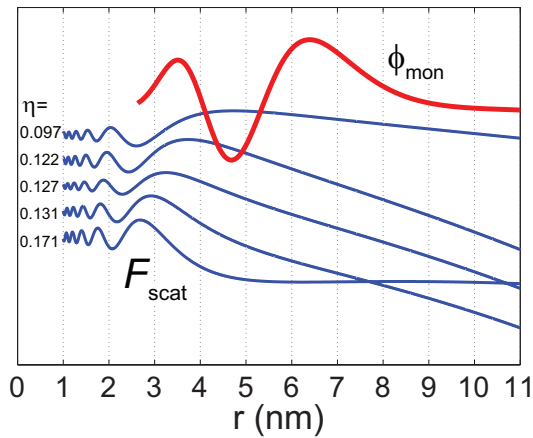}\\
  \caption{ Calculated variation of the scattering wavefunction $F_{\text{scat}}$ for various $\eta$.  A large modification of the wavefunction inside the centrifugal barrier ($r<7.0$~nm) can be seen within a small variation of $\eta$.     }\label{fig4}
\end{figure}

The data with the monitor laser is fitted using an inelastic rate equation $\dot{n}=-2K_{\text{total}}n^2$, where $n$ is the atom density, with a PA rate coefficient $K_{\text{total}}=K_{\text{tun}}+K_{\text{mon}}$.  We fit the atom loss spectra with fitting parameters $\Gamma_{\text{tun}}$ and $K_0$,  with $\gamma/2\pi=364$~kHz used for the molecular state radiative decay rate, and the results are shown in solid lines in Fig.~\ref{fig2}(a) and \ref{fig2}(b).  To include the effect of the collision at different energies, thermal averaging is included in the fit, assuming Maxwell-Boltzmann velocity distribution of the atoms for the given temperature.
The calculated results fit the data remarkably well.  Although the fit slightly overestimates the atom loss near the resonance, the overall fitting quality including the reproducibility of the asymmetric spectra is satisfiable.
We obtained the best fit with $\Gamma_{\text{tun}}/2\pi=16.5$ and 16.2~kHz, where the maximum scattering phase shift $\eta=0.149(2)$ and 0.144(3) radian and corresponding atom loss rate coefficient $K_{\text{tun}}=28.0\times 10^{-12}$ and $17.2\times 10^{-12}$~cm$^3$/s are calculated, for $T=7.2$ and 4.5~$\mu$K, respectively.  From $\Gamma_{\text{tun}}$ obtained from the fit, we also calculate the atom loss spectra with a loss rate coefficient $K_{\text{total}}=K_{\text{tun}}$ for the data without the monitor laser and results are shown in dotted lines in the same figure.  The loss spectra are well reproduced with the fitted parameter.

The calculated variation of $\eta$ from the obtained value of $\Gamma_{\text{tun}}$ at $T=7.2$~$\mu$K is shown in Fig.~\ref{fig2}(c).  The phase shift variation $\delta\eta=\eta_{max}-\eta_0=0.022$~radian is observed.  On the blue side of the resonance, the OFR induced $\eta$ crosses the $\eta_{min}=0.116$, where we expect the Franck-Condon factor for the monitor PA to be zero.  The atom loss curve with the tune and monitor laser (solid line) tangentially connect to the atom loss with the tuning laser only (dotted line) at this region.   From all the data including that for a different temperature, we calculated the average optical volume\cite{Goyal2010} $V_{\text{opt}}=\Gamma /2k^3\gamma =27(6)$~nm$^3$ at the tuning laser intensity $I=1$~W/cm$^2$ for this transition which is lower than the previously calculated value 39.2~nm$^3$ at $I=1$~W/cm$^2$, which is the average value over different projections of the partial-wave angular momentum and the excitation photon helicity \cite{Goyal2010}.  The uncertainties in the measurement are mainly from the PA lasers and atom sample characterizations.  The uncertainty in the laser intensity determination is the dominant source of error, which includes the power fluctuation, beam size uncertainty, and power meter calibration error, resulting in total of 23\% uncertainty.  The atom number fluctuation is approximately 5~\%, while the temperature variation is limited to about 3~\%.    While the values obtained in experiment and theory show a discrepancy, a typical $p$-wave scattering volume (as in van der Waals length scale in the $s$-wave scattering) is on the order of (100~$a_0$)$^3　\sim 150$~nm$^3$, where $a_0$ is Bohr radius.  The observed difference between the experiment and theory is much smaller compared to this typical scale size.

One may wonder why the change of the monitor PA rate is rather strong, showing strong asymmetric atom loss spectra, while the phase shift variation $\delta \eta$ is not significantly large.
In Fig.~\ref{fig4}, we show calculated scattering wavefunctions for $\eta$ ranging from 0.097 to 0.171 radian.
Despite the small variation in $\eta$, a drastic change in the shape of the wavefunction can be observed inside the centrifugal barrier ($r<7.0$~nm).
At $\eta_0$, the scattering wavefunction is away from the shape resonance condition, and the penetration inside the  centrifugal barrier is small and thus the amplitude of the wavefunction inside the centrifugal barrier is small.
When the OFR causes a change of the phase shift $\eta$, even if it is not so large, a large modification of the wavefunction can take place inside to alter the Franck-Condon factor with respect to $\phi_{\text{mon}}$.
By choosing the molecular state which is located well inside the centrifugal barrier, the monitor PA signal amplifies the small shift of $\eta$, resulting in a large spectral modification.

In conclusion, we successfully observed the $p$-wave OFR effect in the vicinity of the previously reported PRL molecular state.  The scattering phase shift variation of $\delta\eta=0.022$ radian at about 5~$\mu $K is observed.  In order to realize a larger phase shift $\eta$ at a low temperature, it is beneficial to use an atomic species which has a $p$-wave shape resonance, which results in an enhancement of $\Gamma_{\text{tun}}$ through the large Frack-Condon factor.
The $^{173}$Yb homonuclear pair has a $p$-wave shape resonance at the collision energy of tens of micro Kelvin \cite{Kitagawa2008}, and thus it would be a good candidate for an efficient $p$-wave OFR.
A future investigation will also include a use of ultra-narrow transitions to $^3P_{0,2}$ states in alkaline-earth-like atoms to further suppress the atom loss and heating of the sample.

We acknowledge useful discussions with B. Deb, I. Deutsch, I. Reichenbach, and P. Zhang.
This work is supported by the Grant-in-Aid for Scientific Research of JSPS (No. 18204035, No. 21102005C01, and No. 21104513A03 Quantum Cybernetics), GCOE Program The Next Generation of Physics, Spun from Universality and Emergence from MEXT of Japan, and World-Leading Innovative R\&D on Science and Technology (FIRST). S. T. and S. S. acknowledge support from JSPS.


\end{document}